\documentclass[epj]{svjour}
\usepackage{graphics}
\usepackage{amsmath}
\usepackage{breqn}
\usepackage{mathrsfs}

\usepackage{mathtools}

\usepackage{caption}
\usepackage{subcaption}
\usepackage{ulem}
\usepackage{soul}
\usepackage{color}
\usepackage{geometry}
\usepackage{url}
\usepackage{ulem}

\def\erf{\text{erf}}
\def\exp{\text{exp}}

\def\ef{\text{eff}}

\def\RA{\Rightarrow}
\def\inf{\infty}

\definecolor{Blue}{RGB}{46,100,254} %Jaime
\definecolor{Cafe}{RGB}{150, 75, 0} %Tom
\definecolor{Black}{rgb}{0., 0., 0.}
\definecolor{ol}{RGB}{107,142,35}
\definecolor{v}{rgb}{0.6, 0.2, 0.8} %VM

\graphicspath{{figures_summary_pantheon/}} % Location of the graphics files

\begin{document}

\title{Constraints on barotropic dark energy models by a new phenomenological $q(z)$ parameterization.}
\author{Jaime Rom\'an-Garza\inst{1,2},  Tom\'as Verdugo\inst{2}, Juan Maga\~na\inst{3} \and Ver\'onica Motta\inst{3}
}                     % Do not remove
\institute{Facultad de Ciencias F\'isico Matem\'aticas, Universidad Aut\'onoma de Nuevo Le\'on, San Nicol\'as de los Garza, M\'exico \and Instituto de Astronom\'ia, Universidad Nacional Aut\'onoma de M\'exico, Apdo. postal 106, CP 22800, Ensenada, B.C, M\'exico \and Instituto de F\'isica y Astronom\'ia, Universidad de Valpara\'iso, Avenida Gran Breta\~na 1111, Valpara\'iso, Chile}
\date{Received: 05-Jul-2019 / Revised version: 09-Oct-2019}
\abstract{
In this paper, we propose a new phenomenological two parameter parameterization of $q(z)$  to constrain barotropic dark energy models by considering a spatially flat Universe, neglecting the radiation component, and reconstructing the effective equation of state (EoS). This two free-parameter EoS reconstruction shows a non-monotonic behavior, pointing to a more general fitting for the scalar field models, like thawing and freezing models. We constrain the $q(z)$ free parameters using the observational data of the Hubble parameter obtained from cosmic chronometers, the joint-light-analysis (JLA) Type Ia Supernovae  (SNIa) sample, the Pantheon  (SNIa) sample, and a joint analysis from these data.  We obtain, for the joint analysis with the Pantheon (SNIa) sample a value of $q(z)$ today, $q_0=-0.51\substack{+0.09 \\ -0.10}$, and a transition redshift, $z_t=0.65\substack{+0.19 \\ -0.17}$ (when the Universe change from an decelerated phase to an accelerated one). The effective EoS reconstruction and the $\omega'$-$\omega$ plane analysis  point towards {a transition over the phantom divide, i.e. $\omega =-1$}, which is consistent with a non parametric EoS reconstruction reported by other authors.
\PACS{
      {98.80.-k}{Cosmology}   \and
      {95.36.-x}{Dark energy}
     } % end of PACS codes
} %end of abstract
\authorrunning{Rom\'an-Garza, Verdugo, Maga\~na, Motta} \titlerunning{Constraints on DE models by a new $q(z)$.}
\maketitle

\section{Introduction}
\label{sec:S1}

Several cosmological observations indicate that the Universe  experiments a late-time acceleration \cite{Planck:2015XIII}. This feature was evidenced for the first time by the observations of distant Type Ia Supernovae (SNIa) \cite{Riess:1998,Perlmutter:1999} and is one of the major puzzles in modern cosmology. In general, there are two ways to explain this mysterious cosmic phase: i) to postulate a fluid with negative pressure, the so-called dark energy (DE), into the canonical  Einstein's  general relativity theory or ii) to modify the gravity laws. Between these two approaches, numerous models have been proposed. Most of them can explain a wide range of the cosmological observations and distinguishing among them is not a trivial problem. Despite of this, one simple model  has been established as the standard, the one with a cosmological constant associated to the quantum vacuum fluctuations $\Lambda$ with cold dark matter ($\Lambda$CDM). Nevertheless, it has theoretical problems \cite{Copeland:2006,Bamba:2012cp} which motivates further studies of 
 alternative models \cite{Li:2011sd}. For instance, some of those consider a dynamical DE involving scalar fields, 
like quintessence \cite{Wetterich:1988,Peebles:1988,Caldwell:1998}, phantom \cite{Caldwell:2002,Chiba:2000}, quintom \cite{Guo:2005}, and k-essence fields \cite{ArmendarizPicon:2000dh,ArmendarizPicon:2000ah}.  An advantage of these models is that the DE equation of state (EoS) evolves with time, and thus it can be parameterized by a function of the scale factor (redshift, as proposed by \cite{Chevallier:2000qy,Linder:2003}) to explore its cosmological behavior.

The standard way to examine these models is to calculate the Friedmann and Raychaudhuri equations in a background cosmology to constrain their free parameters (see for example \cite{Pantazis:2016}). A model-independent approach is to investigate the cosmographic parameters that characterize the kinematics of the cosmic expansion \cite{luongo2011cosmography,Aviles:2012ay,padeaprox,Demianski:2012,gruber2014cosmographic,Zhang:2017EPJC} . The advantage of this procedure is that the only assumption is the Cosmological Principle, i.e. an homogeneous and isotropic Universe, without speculating about its composition. Indeed, it is very common to consider the Hubble parameter, $H\equiv \dot{a}/a$, and the deceleration parameter, $q(a)\equiv -\ddot{a}a/\dot{a}^{2}$  \footnote{Alan Sandage claimed that the cosmic expansion can be determined by these two parameters at z=0 \cite{Sandage1970}}. However, higher order derivatives of the scale factor $a$, such as jerk and snap, can be also considered, e.g. \cite{Mamon:15}. By probing the cosmographic parameters using cosmological data, we can associate them to a given dynamical DE entity and reconstruct its features as well as the Universe dynamics. In this vein, several authors have proposed a number of functions to parameterize the deceleration parameter $q(z)$ (see for example \cite{delCampo:2012ya,Nair:2011tg,Santos:2010gp,al2016divergence,Mamon:2016dlv,Mamon:15} for recent studies) and associate its features to a some DE model.

\begin{figure*}
\centering
\captionsetup{justification=centering}
\resizebox{0.49\textwidth}{!}{\includegraphics{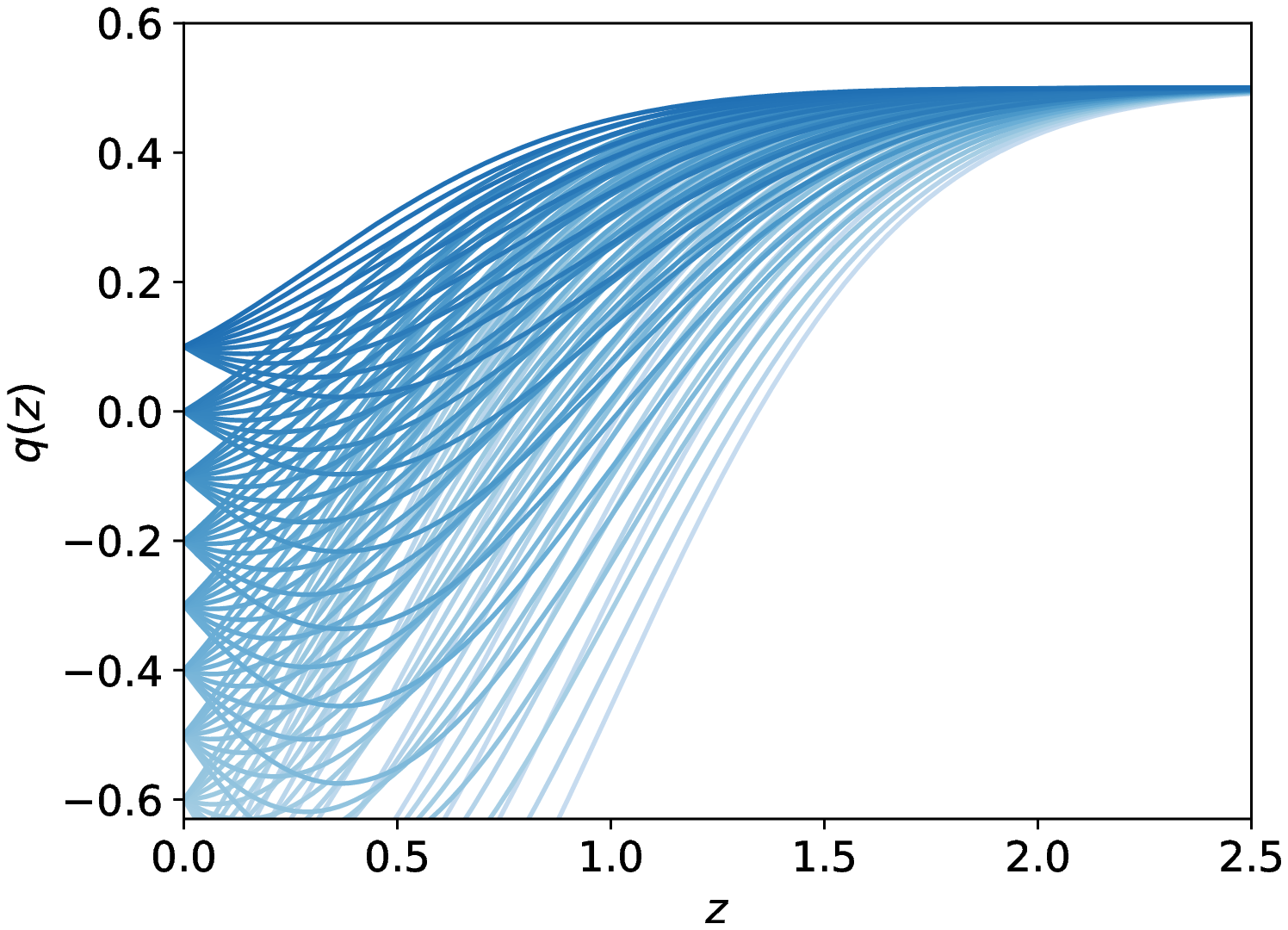}}
\resizebox{0.49\textwidth}{!}{\includegraphics{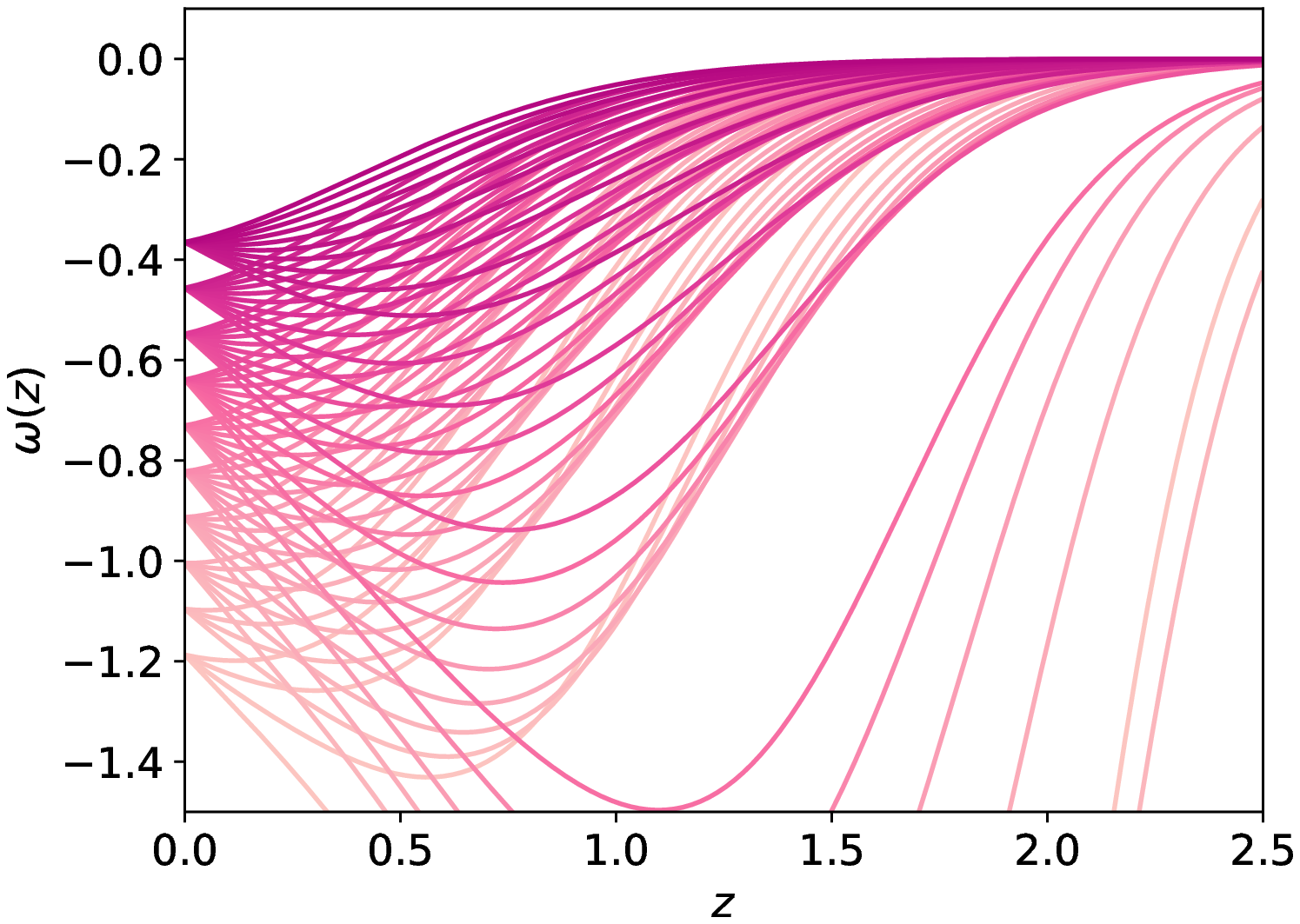}}
\caption{\textit{Left panel.-} Functional form of the proposed $q(z)$ given by equation \eqref{eq:q} for different $(q_{0}$,$z_{c})$ values. Notice that both an accelerated and decelerated stage at $z=0$ are allowed. \textit{Right panel.-} Functional form of $\omega(z)$ calculated through equation\eqref{eq:wdef}}
\label{fig:Qz_sig}
\end{figure*}
%%%%%%%%%%%%%%%%%%%%%%%

The motivation of the present work is to propose a new parameterization of the deceleration parameter as function of redshift, based only in the cosmological principle, and able to generate an EoS which describes both, slowly and rapidly behaviors \cite{Bassett:2004ApJ}.  The ansatz is a continuous and differentiable function  that is valid from the matter domination epoch until the near future. We constrain the $q(z)$ free parameter by performing a Bayesian analysis for which we employ the latest compilation of observational Hubble data (hereinafter OHD) from cosmic chronometers and Type Ia Supernova. Using the mean value parameters, we reconstruct an effective EoS to  the dynamical dark energy.

The paper is organized as follows. In sec. \ref{sec:S2} we state the theoretical framework and present the parametric equation of the deceleration parameter. Section \ref{sec:S3} provides a description of the data sets and the methodology used to constrain the parameters of the deceleration parameter. The sec. \ref{sec:S4} presents the obtained EoS, and the tools to discriminate between different DE models. Finally, in sec. \ref{sec:S6} the remarks and conclusions are presented.

\section{Theoretical framework}
\label{sec:S2}
\subsection{Proposed parameterization for the deceleration parameter}

\noindent The deceleration parameter as function of $H(z)$ is  \\
\begin{equation}
\label{eq:qdef}
q=-\left(1+ {\dot{H}\over{H^2}} \right)\text{,}
\end{equation}
if $q>0$ the Universe is at a decelerated phase, otherwise $q<0$ corresponds to an accelerated phase. By integrating the eq. \eqref{eq:qdef}, the Hubble parameter can be written as:\\
\begin{equation}
\label{eq:Hdef}
H(z)=H_0\ \text{exp}\left( \int_0^z {{1+q(z')}\over{1+z'}} dz' \right),
\end{equation}
where $H_0$ is the Hubble parameter at the present epoch and 
$z=(1/a) -1$ is the redshift.\\

The OHD suggest that $q<0$ at the present epoch and $q>0$ during an early epoch when the matter dominates as shown in ref.~\cite{Turner:2001mx,Bautista:2017zgn}. The structure formation at this early epoch is explained by a decelerated phase, so the value of the deceleration parameter transit from positive in the past to negative at the present. The parameterization of the deceleration parameter is a useful method to reconstruct cosmological parameters and constrain the dynamical evolution of the universe in a general scheme ~\cite{Mamon:2017rri}. There are several parameterizations for $q(z)$ reported in the literature, see refs.~\cite{Turner:2001mx,Mamon:2017rri,Cunha:2008abc,Cunha:2008mt,delCampo:2012ya,Nair:2011tg,Santos:2010gp,Gong:2006tx,xu2009cosmic,al2016divergence,Mamon:2016dlv,Xu:2007nj,Mamon:15}. We propose a new one as follows

\begin{equation}
\label{eq:q}
q(z)=q_1+(q_0-q_1)(z+1)e^{z_c^2-(z+z_c)^2},
\end{equation}

\noindent where, $q_0$ and $q_1$ are the values  for the deceleration parameter at the present epoch, and at high redshift, respectively. We set $q_1$ = 0.5 to consider the matter-dominated epoch of the Universe. The characteristic redshift, $z_c$, is a free parameter  related to the transition redshift, $z_t$, the redshift at which the Universe underwent a transition from deceleration to an acceleration phase. This is a well behaved parameterization (see figure~\ref{fig:Qz_sig}) that can reproduce a soft step transition, as well as changes in concavity in the deceleration parameter (notice that both an accelerated and decelerated stage at $z=0$ are allowed), and facilitates the analytical reconstruction of other cosmological parameters like $H(z)$, and $w(z)$. Note how combinations of $q_0$ and $z_c$ can yield the same transition redshift.  \\

Substituting the equation \eqref{eq:q} into the equation \eqref{eq:Hdef}, we obtain the analytical expression  for the Hubble parameter in terms of $z$:
\begin{equation}
\label{eq:H}
H(z)=H_0(z+1)^{q_1+1}e^{\xi \eta},  
\end{equation}

\noindent where $\xi={(\sqrt{\pi}/2)}(q_0-q_1)e^{z_c^2}$, $\eta = \erf(z+z_c)-\erf(z_c)$,  
and $\erf(x)$ is the error function of $x$. This is the expression that is fitted to the data.

\subsection{The effective Equation of State}

With the metric for a spatially flat Friedmann-Lema\^{i}tre-Robertson-Walker (FLRW) space-time,

\begin{equation}
\label{eq:metric}
ds^2=-dt^2+a^2(t)\{ dr^2+r^2d\Omega^2 \},
\end{equation}

\noindent and considering a space-time composed of a non-relativistic component $\rho_{m}$ and a barotropic fluid with an effective density $\rho_\ef$ and an effective pressure $p_\ef$, the Einstein field equations in units of $8\pi G=c=1$ are obtained following ref.~\cite{Copeland:2006} as 
%and considering a space-time filled with a scalar field $\phi$, having a potential $\Vph$, the Einstein field equations are obtained as in ref.~\cite{Mamon:15} as
\begin{align}
\label{eq:Eqe1}
3H^2 &= \rho_m+\rho_\ef\text{,}\\
\label{eq:Eqe2}
2\dot{H}+3H^2 &= -p_\ef\text{,}
\end{align} 
and the effective EoS is written as 
\begin{align}
\label{eq:defw}
\omega=\frac{p_\ef}{\rho_\ef}\text{.}
\end{align} \\
Substituting \eqref{eq:Eqe1} and \eqref{eq:Eqe2} in \eqref{eq:defw}, the EoS in terms of $q(z)$ and $H(z)$ is obtained  following ref.~\cite{Mamon:2016dlv} as 
\begin{align}
\label{eq:wdef}
\omega(z) = \frac{2}{3}{{q(z)-{1 \over 2}} \over {1-\Omega_{m,0}(1+z)^3 \left({ H_0 \over {H(z)}} \right) ^2}}\text{.}
\end{align} \\

\noindent where $\Omega_{m,0}$ is the matter density parameter $\Omega_{m}=\rho_{m}/\rho_{crit}$ evaluated at $z=0$\footnote{Here $\rho_{crit}$ is the standard critical density defined as $3H(z)^{2}/8\pi G$}.

By substituting equations \eqref{eq:q} and \eqref{eq:H} in equation \eqref{eq:wdef}, we obtain the expression

\begin{multline}
\label{eq:A_w}
\omega(z)={\frac{2}{3}} \times\\
{\frac{(q_0-q_1)(z+1)\text{exp}(z_c^2-(z+z_c)^2)}{1-\Omega_{m,0}(1+z)^{1-q_1}\text{exp}(-2\xi \eta)}} \text{.}
\end{multline}

\noindent  The right panel of Figure~\ref{fig:Qz_sig}  shows how the EoS changes for different values of $q_0$ and $z_c$.  The reconstruction of $\omega(z)$  yield distinct DE behaviors when the barotropic fluid is associated to a minimally coupled scalar field: quintessence ($-1 \le \omega(z)  \le 1$), phantom ($\omega(z)<-1$) or {{even crossing the phantom divide, $\omega=-1$, e.g. quintom models (where the DE component moves across the quintessence and phantom regions through two scalar fields) see \cite{Copeland:2006} and references therein.}}  In contrast to some  $\omega(z)$ parameterizations analyzed in the literature \cite{Pantazis:2016,Chevallier:2000qy,Linder:2003,Jassal:2004ej,Feng:2012gf,Sendra:2012,Rezaei:2017yyj,Barboza:2008rh}, our  EoS concavity changes from low to high $z$ values if there is at least one inflexion point at $z>0$. Some authors have proposed a more general form for the EoS parameterization, with a different approach in which a transition function introduces a rapid evolution of $w(z)$ \cite{Bassett:2004ApJ,Corasaniti:2003PhRvD,Corasaniti:2004PhRvD}. In the present work, we obtain a similar result, however the main difference is that the behavior of the EoS is a direct result of the proposed $q(z)$ parameterization. Indeed, this further highlights the importance of the proposed functional form for the deceleration parameter.

\section{Observational data and methodology}
\label{sec:S3}
In this section we introduce the cosmological data and the methodology used to constraint the $q(z)$ free parameters of the equation~\eqref{eq:q}.

\subsection{Observational Hubble Data from cosmic chronometers.}
Several authors have shown that the OHD can be used to constrain cosmological parameters. There are two techniques to
measure the cosmic expansion at different redshifts: using the baryon acoustic oscillation analysis or applying the differential age technique (DA) in cosmic chronometers, i.e. passive-early-type galaxies. This last method was proposed by \cite{Jimenez:2001gg} and measures $H(z)$ using the following relation for two early-type galaxies separated by a small redshift interval $\Delta z$ 
\begin{equation}
H(z)=-\frac{1}{1+z}\frac{\mathrm{dz}}{\mathrm{dt}},
\end{equation}
where $\mathrm{dz/dt}$  is measured by estimating the differential age $\Delta{t}$  with the $4000 \mbox{\AA}$ break ($D4000$) feature in their spectra.

%\resizebox{0.45\textwidth}{!}
%%%%%%%%%%%%%%%%%%%%%%%%%%%%%% Image %%%%%%%%%%%%%%%%%%%%%%%%%%%%%%%%%%%%%%
%%%%%%%%%%%%%
\begin{figure*}
\centering
%\begin{figure}
\begin{center}\vspace{ 0.25cm}
\captionsetup{justification=centering}
%\resizebox{0.45\textwidth}{!}{\includegraphics{contours_joint_constraints}}
\resizebox{0.80\textwidth}{!}{\includegraphics{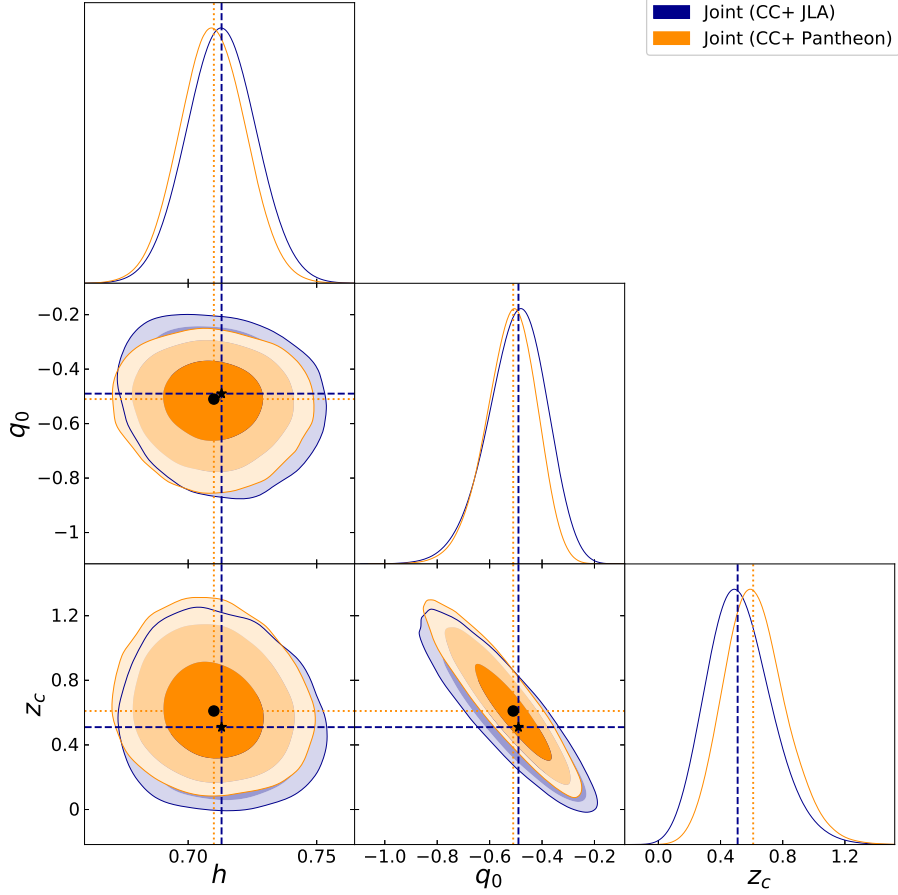}}
\captionof{figure}{1D marginalized posterior distributions and the 2D $68\%$, $95\%$, and $99.7\%$ confidence levels for the $h$, $q_{0}$ and $z_{c}$ parameters for the joint analysis of SNIa+CC. The circle and the star represent the parameter mean values for the CC+JLA and the CC+Pantheon samples, respectively. \label{fig:confidence_contours}}
\end{center}\vspace{1.cm}
\end{figure*}

We employ the latest OHD obtained from DA in cosmic chronometers, which contains $31$ data points covering $0 < z < 1.97$, compiled by \cite{Juan:2017} and references therein.
The figure-of-merit for the OHD is written as
\begin{equation}
\chi_{\mathrm{OHD}}^2 = \sum_{i=1}^{31} \frac{ \left[ H(z_{i}) -H_{DA}(z_{i})\right]^2 }{\sigma_{H_i}^{2}},\label{fOHD}
\end{equation}
where $H(z_{i})$ is the theoretical Hubble parameter,
$H_{DA}(z_{i})$ is the observational one at redshift $z_{i}$, and
$\sigma_{H_i}$ is its uncertainty. 

%\resizebox{0.45\textwidth}{!}
%%%%%%%%%%%%%%%%%%%%%%%% Image %%%%%%%%%%%%%%%%%%%%%%%%%%%%%%%%

%%%%%%%%%%%%%%%%%%%%%%%%%%%%%
\begin{figure*}
\centering
\captionsetup{justification=centering}
\resizebox{0.49\textwidth}{!}{\includegraphics{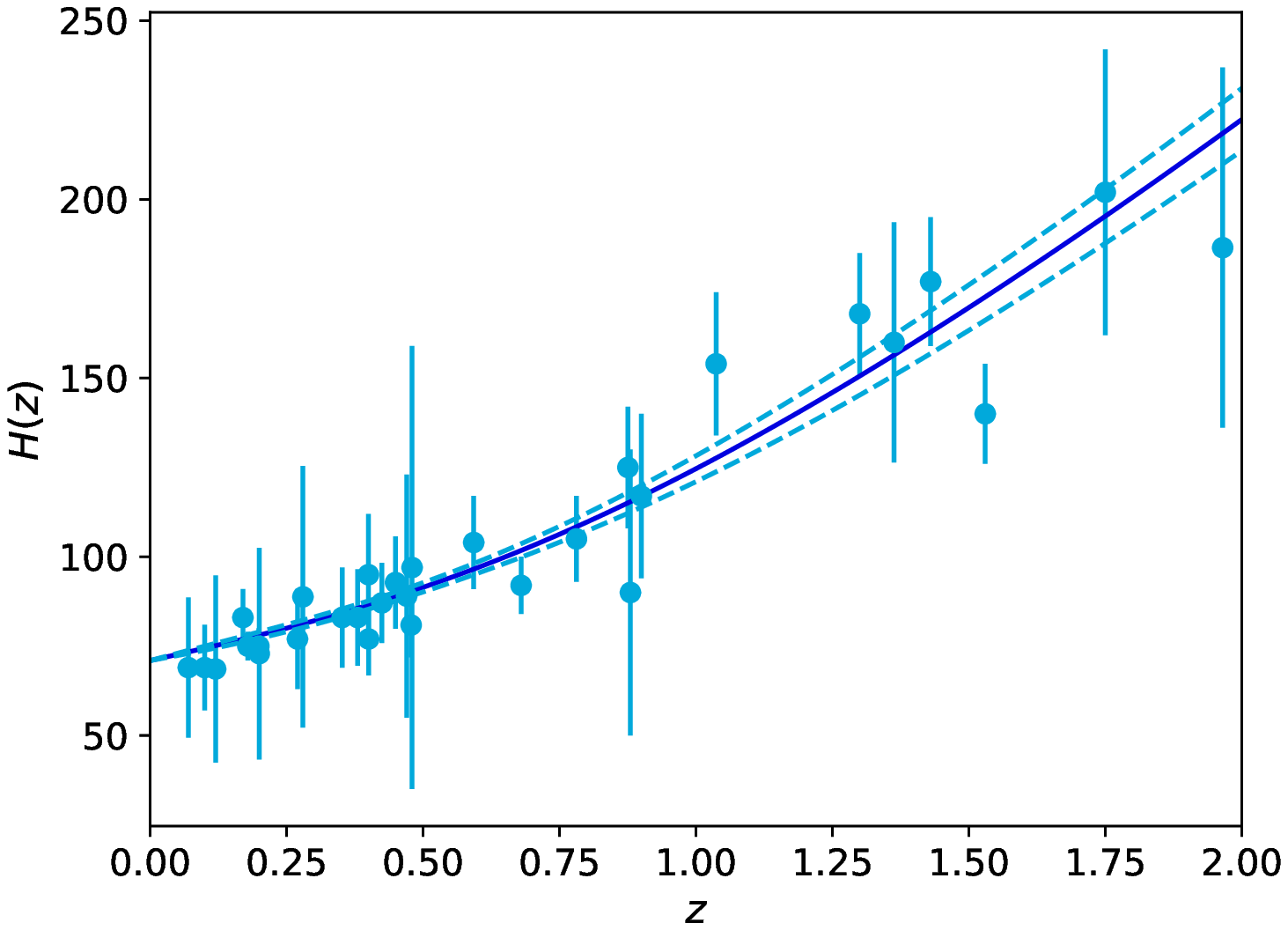}}
\resizebox{0.49\textwidth}{!}{\includegraphics{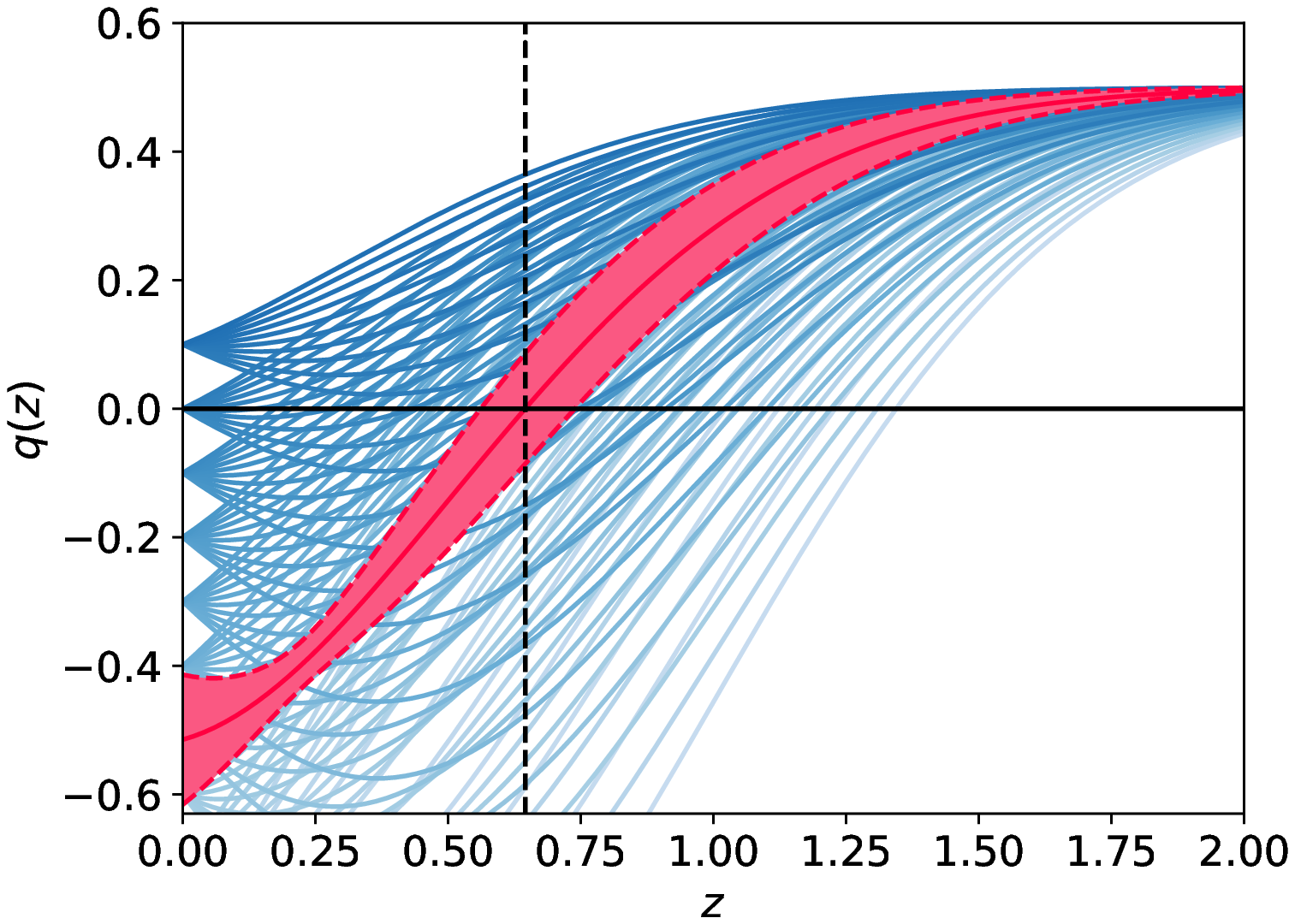}}
\captionof{figure}{Fit to OHD (left panel) and the reconstructed $q(z)$ (right panel) using the joint analysis (CC+Pantheon) constraints. The dashed and red shadow regions show the $1\sigma$ confidence limits estimated from a MCMC analysis. The black dashed line in the right panel represents the transitional redshift, $z_t$, for the joint analysis mean value.\label{fig:hz_qz}}
\end{figure*}
%%%%%%%%%%%%%%%%%%%%%%%%%%%%%

\subsection{Type Ia Supernovae }
The standard test to investigate the accelerating expansion is employing the observations of type SNIa at high redshifts. 
We use two of the latest SNIa compilations, the so-called joint-light-curve-analysis (JLA) \cite{Betoule:2014} sample, that contains $740$ points spanning a redshift range $0.01<z<1.2$, and the Pantheon sample \cite{Scolnic:2018} containing $1048$ points in the redshift range $0.001<z<2.3$.

\subsubsection{JLA SNIa sample}
The figure-of-merit for the JLA data is given by

\begin{equation}
\chi^{2}_{\mbox{JLA}}=\left(\hat{\mu}_{\mathrm{JLA}} - \mu_{qz}\right)^{\dag}\mathrm{C_{\eta}^{-1}}\left( \hat{\mu}_{\mathrm{JLA}} - \mu_{qz} \right), \label{fJLA}
\end{equation}
where  $ \mu_{qz} = 5 \log_{10} \left( d_L / 10 \mathrm{pc} \right)$ is the theoretical distance modulus for the $q(z)$ parameterization and $d_{L}$ is the luminosity distance given by
 \begin{equation}
d_{L}=c(1+z) \int^{z}_{0}\frac{\mathrm{dz}^{\prime}}{H(z^{\prime})}.
\label{eq:dl}
\end{equation}
The observational distance modulus, $\hat{\mu}$, for the the JLA data reads as
\begin{equation}
\hat{\mu}_{\mathrm{JLA}} = m_{b}^{\star} - \left( M_{B} -  \alpha \times X_{1}   +  \beta \times C  \right),  \label{muhat}
\end{equation}
where $m_{b}^{\star}$ corresponds to the observed peak magnitude, $M_{B}$ is the $B$-band absolute magnitude. The $X_{1}$ and $C$ variables describe the time stretching of the light-curve and the Supernova color at maximum brightness respectively. The $\alpha$, and $ \beta $ coefficients are nuisance parameters. For JLA sample, the absolute magnitude $M_{B}$ is related to the host stellar mass, $M_{stellar}$ by the step function:
\begin{align}
M_{B} = \left\{ \begin{array}{cc} 
                M_{B}^{1} & \hspace{5mm}  \rm{if} \ M_{stellar} < 10^{10} M_\odot\ ,\\
              M_{B}^{1} + \Delta_{M} & \hspace{5mm} \rm{otherwise.} \\
                \end{array} \right.
\end{align}
Finally, $\mathrm{C_{\eta}}$ is the covariance matrix\footnote{available  at \\  \scriptsize{\url{http://supernovae.in2p3.fr/sdss_snls_jla/ReadMe.html}}} of $\hat{\mu}$ provided by  \cite{Betoule:2014}, which takes into account several statistical and systematic errors in the SNIa data.

\subsubsection{Pantheon SNIa sample}
The observational distance modulus $\mu_{\mathrm{PAN}}$ for Pantheon SNIa can be measured as

\begin{equation}
  \mu_{\mathrm{PAN}} = m_{b}^{\star} - M_{B} + \alpha \times X_{1} - \beta \times C + \Delta_M+ \Delta_B, 
  \label{eq:mu}
\end{equation}
%
%where $M$ is the absolute $B$-band magnitude of a %fiducial SNIa with $x_1 = 0$ and $c = 0$, 
where the parameters  $m_{b}^{\star}$, $M_{B}$, $\alpha$, $X_{1}$, $\beta$, and $C$ are the same as the JLA sample.
$\Delta_M$ is a distance correction based on the host-galaxy mass of the SNIa and $\Delta_B$ is a distance correction based on predicted biases from simulations.  
It is worthy to note that \cite{Scolnic:2018} provided 
$\tilde{\mu_{\mathrm{PAN}}}=\mu_{\mathrm{PAN}}+M_{B}$, thus, we can marginalize over the $M_{B}$ parameter. The marginalized figure-of-merit for the Pantheon sample is given by
\begin{equation}
\chi_{Pan_{Mmarg}}^{2}=a +\log \left( \frac{e}{2\pi} \right)-\frac{b^{2}}{e}, \label{fPan}
\end{equation}
\noindent
where $a=\Delta\boldsymbol{\tilde{\mu}}^{T}\cdot\mathbf{C_{P}^{-1}}\cdot\Delta\boldsymbol{\tilde{\mu}},\, b=\Delta\boldsymbol{\tilde{\mu}}^{T}\cdot\mathbf{C_{P}^{-1}}\cdot\Delta\mathbf{1}$,\, $e=\Delta\mathbf{1}^{T}\cdot\mathbf{C_{P}^{-1}}\cdot\Delta\mathbf{1}$, and $\Delta\boldsymbol{\tilde{\mu}}$ is the vector of residuals between the model distance modulus and the observed $\tilde{\mu_{\mathrm{PAN}}}$. The covariance matrix $\mathbf{C_{P}}$ can be constructed as $\mathbf{C_{P}}=\mathbf{C_{P,stat}}+\mathbf{C_{P,sys}}$, where $\mathbf{C_{P,sys}}$ is the systematic covariance matrix, and $\mathbf{C_{P,stat}}$ is a diagonal matrix which contains the statistical errors on $\tilde{\mu_{\mathrm{PAN}}}$. We refer the interested reader to \cite{Scolnic:2018} for a detailed description how these matrices are constructed.

\subsection{Fitting the data}
%\label{sec:S5}

To estimate the values of the parameters $q_0$ and $z_c$  from  equation~\eqref{eq:q},  a Markov Chain Monte Carlo (MCMC) Bayesian statistical analysis is performed using the Affine-Invariant MCMC Ensemble sampler from the emcee
Python module~\cite{Foreman:2013}. We perform the following cases: using only the CC data, only a SNIa sample (JLA or Pantheon), and the joint analysis CC + SNIa (JLA or Pantheon).The computations are running with $1500$ steps to stabilize the estimations
(burn-in phase), and $5000$ MCMC steps using $600$ walkers. We assume the following flat priors for all cases: $h$ $\in$ $[0,1]$, $q_{0}$ $\in$  $[-1,1]$, $z_{c}$ $\in$ $[0,2]$. When the JLA sample is used in the analysis, we also consider  $ M^{1}_{b}$ $\in$ $[-20.0$ $,-18.0]$, $\Delta_{M} \in [-0.1,0.1]$, $\alpha \in [0.0, 0.2]$, and $\beta \in [0.0, 4.0]$. To assess the convergence of our analysis, a Gelman-Rubin test is employed. 

We assume a Gaussian likelihood when the parameter estimation is performed using only a data set.
The goodness of the fit for the joint analysis is quantified by a total $\chi^2$ defined as:

\begin{equation}\label{eq:Chi2}
\chi^2_{T} = \chi^2_{\mathrm{OHD}} + \chi^2_{\mathrm{SNIa}},
\end{equation}

%%%%%%%%%%%%%%%%%%%%%%%%%%%%%%%5 Image %%%%%%%%%%%%%%%%%%%%%%%%%%%%%%%%
%%%%%%%%%%%%%%%%%%%%%%%%%%%%%%%%%
\begin{figure}
\centering
\captionsetup{justification=centering}
\resizebox{0.52\textwidth}{!}{\includegraphics{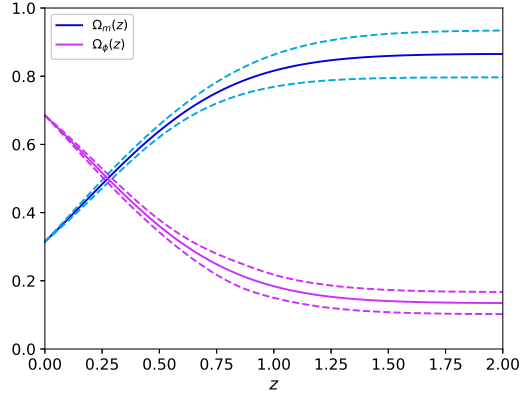}}
\captionof{figure}{The matter and DE density parameter of the Universe using the joint analysis mean values. The dashed regions show the $1\sigma$ confidence limits estimated from a MCMC analysis.
 \label{fig:dw}}
\end{figure}
%%%%%%%%%%%%%%%%%%%%%%%%%%%%%%%%%%

\noindent where $\chi^2_{\mathrm{OHD}}$ is calculated using eq. \eqref{fOHD}. And $\chi^2_{\mathrm{SNIa}}$ is calculated using eq. \eqref{fJLA} or eq. \eqref{fPan} for the JLA or Pantheon sample respectively. Thus, a joint Gaussian likelihood can be expressed as:

\begin{equation}\label{eq:likeli}
\mathcal{L}_{\mathrm{joint}} \propto  \exp(-\chi^2_{T}/2),
\end{equation}

\noindent where $\mathcal{L}_{\mathrm{joint}}$ is the product of the likelihood functions of each data set.  

The mean values of the fits are presented in Table~\ref{tab:par}. Figure~\ref{fig:confidence_contours} shows the confidence contours obtained for the joint analysis, for both, the JLA and the Pantheon samples. In the left panel of figure~\ref{fig:hz_qz} we show the OHD along with the function given by equation \eqref{eq:H} using the  mean values obtained from the joint analysis (CC + Pantheon) fitting. In the right panel of the same figure is the reconstructed deceleration parameter with these same constraints.

Along with the narrow constraints obtained with the joint analysis (see figure \ref{fig:confidence_contours}), we note an anticorrelation between  $z_{c}$ and $q_{0}$ parameters. {{This degeneracy has a mathematical origin: as $q_{0}$ becomes less negative, the transition redshift $z_t$ is larger\footnote{The parameter $z_t$ is obtained solving the expression $0=0.5+(q_0-0.5)(z_t+1)e^{z_c^2-(z_t+z_c)^2}$} , which in turn decreases $z_{c}$ (see figure 1).  The $q_{0}$-$z_{c}$ contours at 3$\sigma$  restrict the possible values of the transition redshift approximately between 0.5 and 1.0,  reproducing an accelerated cosmic phase at late times. Additionally, notice that the principal axes of the $h$-$q_{0}$ and $h$-$z_{c}$ confidence contours are parallel to the coordinates axes, indicating that these parameter pairs are uncorrelated. For the case of a spatially flat universe filled with a barotropic fluid (or more than one), it has been shown that $q_{0}$ only depends on the density parameters, and the corresponding EoS of the fluid \cite{luongo2011cosmography,Aviles:2012ay,gruber2014cosmographic,Demianski:2012,Lima:2012bx,Bolotin:2015dja}. This characteristic is also observed in our $h$-$q_{0}$ constraints, which further supports the proposed functional form for the deceleration parameter and the assumption of a barotropic fluid. A similar analysis can be obtained for the  transition redshift $z_{t}$ (related to the $z_{c}$ parameter) in the sense that such parameter is associated with the density parameters of the Universe's components. The result for the $h$-$z_{c}$ constraints depicted in Figure 2 shows also a lack of correlation between both parameters,  reinforcing the proposed model in this work.}

{It is worth to note that the joint confidence contours using the Pantheon sample are slightly  narrower  than  those  obtained with the JLA sample. This feature is also present in the 1D histograms, those estimated with the Pantheon sample are slightly more tight (see also errors in Table~\ref{tab:par}). This is related to several systematic uncertainties in the measurements of the SNIa (e.g., photometry, and astrometry calibration, SN modeling, Milky Way extinction model), see \cite{Scolnic:2018}.} {Our results are consistent with those of \cite{Scolnic:2018}, i.e. Pantheon sample seems to provide tighter cosmological constraints than the JLA sample, although the difference is not statistically significant.}}

A numerical analysis of the roots of $q(z)$ allows to estimate the value of the transition redshift, $z_t=0.65\substack{+0.19 \\ -0.17}$, for the joint data set using the Pantheon sample (we will only make use the results of this joint analysis hereinafter).  This result is consistent with values reported in literature \cite{Cunha:2008mt,Riess:2004nr,Xu:2007nj,Mamon2017,Mamon:15,Rani:2015lia,ishida2008did,dos2016constraining,rani2015transition}, indicating that the Universe passed from a decelerated phase to an accelerated one at $z \approx 0.7$.  The right panel of figure \ref{fig:hz_qz} illustrates the reconstructed $q(z)$ for the joint analysis constraints. Note that  $q_0=-0.51\substack{+0.09 \\ -0.10}$, and the reconstructed $\Omega_m$(z) are in agreement with the dynamics of the standard cosmological model, as well with \cite{aviles2012cosmography,aviles2017toward,gruber2014cosmographic,luongo2011cosmography,giostri2012cosmic,vitenti2015general,davari2018new}. The matter component is dominant with respect to the dark energy component for high redshift values, the opposite occurs at late times (see figure \ref{fig:dw}).

%%%%%%%%%%%%%%%%%%%%%%%%%%%%%%%%%%% Image %%%%%%%%%%%%%%%%%%%%%%%%%%%%%%
%%%%%%%%%%%%%%%%%%%%%%%%%%%%%%%%
\begin{figure*}[h!]\begin{center}
\captionsetup{justification=centering}
\resizebox{0.49\textwidth}{!}{\includegraphics{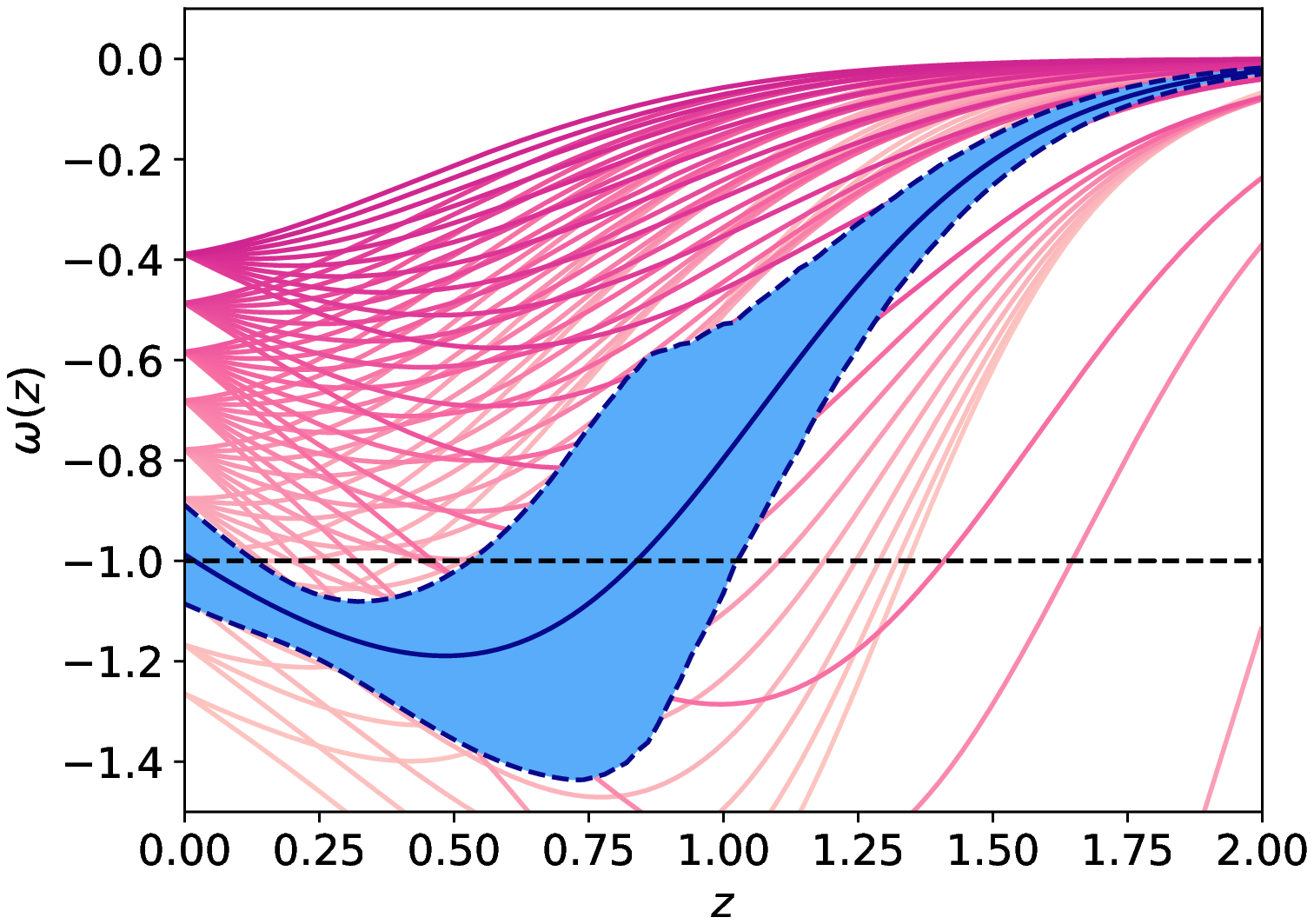}}
\resizebox{0.49\textwidth}{!}{\includegraphics{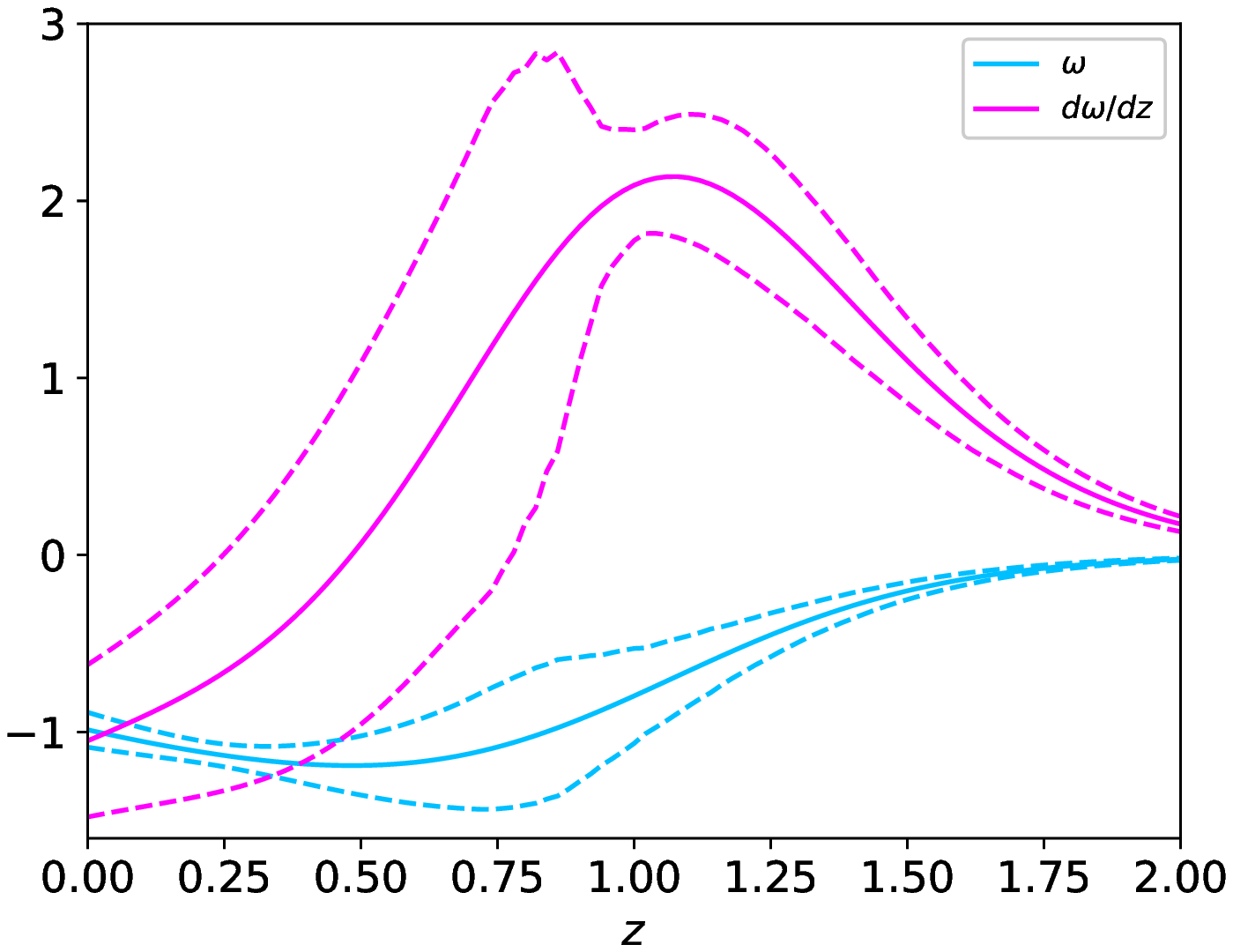}}\\
\captionof{figure}{The $\omega$ reconstruction using the joint analysis (CC + Pantheon) mean values and its functional form (left panel). The horizontal black dashed line represent the phantom divide. The right panel shows the effective EoS (cyan) and its derivative with respect to $z$ (magenta). The regions delimited by dashed lines show the $1\sigma$ confidence limits estimated from a MCMC analysis.\label{fig:param}}
\end{center}\end{figure*}
%%%%%%%%%%%%%%%%%%%%%%%%%%%%%%%%

\section{Dynamical Dark Energy}\label{sec:S4}

\subsection{The resulting EoS}

The left panel in figure~\ref{fig:param} presents the EoS constructed from the equation~\eqref{eq:A_w} using the parameter mean values and $\Omega_{m,0}$ = 0.31 \cite{Aghanim:2018}

%\begin{multline}
%\label{eq:w_final}
%\omega(z)={\frac{2}{3}}\times\\
%(-0.48^{+0.10}_{-0.11}-0.5)(z+1)\text{exp}((0.50^{+0.20}_{-0.19})^2-(z+0.50^{+0.20}_{-0.19})^2)\\
%{\frac{1}{1-0.3(1+z)^{0.5}\text{exp}(-2\xi(\erf(z+0.50^{+0.20}_{-0.19})-\erf(0.50^{+0.20}_{-0.19})))}} \text{.}
%\end{multline}

\begin{equation}
\label{eq:w_final}
\omega(z)=\mathcal{A}(z)\times {\frac{1}{1-\mathcal{B}(z)}} \text{.}
\end{equation}

\noindent where $\mathcal{A}(z)$ is a function of $z$, and  $\mathcal{B}(z)$ could be expressed as

\begin{multline}
\label{eq:w_final_den}
\mathcal{B}(z) = 0.31(1+z)^{0.5} \times \cr  e^{(-2\xi(\erf(z+0.50^{+0.20}_{-0.19})-\erf(0.50^{+0.20}_{-0.19})))}\text{.}  
\end{multline}

\noindent  Although equation~\eqref{eq:w_final} is a well-behaved function, from equation~\eqref{eq:w_final_den} is clear that the denominator may be zero, leading to a singularity in the EoS (see next section). A way to overcome this problem is studying the derivative of the  EoS~\cite{Zhang:2017EPJC}. From eq.~\eqref{eq:wdef}, it is straightforward to show that

\begin{multline}
\label{eq:A_dw}
\frac{\mathrm{d}\omega(z)}{\mathrm{dz}} = \frac{2}{3}\ (1-\Omega_m(z))  (q(z) ( \frac{1}{z+1}-   \cr  2(z+z_c))  q_1(2(z+z_c)-\frac{1}{1+z}))  + \cr 3(q(z)-\frac{1}{2}){\Omega_m(z)}  \left({(1+z)(1-\Omega_m(z))^2}\right)^{-1} \text{.}
\end{multline}

\noindent The equation $\omega$ and the derivative $d\omega/dz$ are shown in the right panel of figure~\ref{fig:param}. The value of the EoS today, $\omega(z)\vert_{z=0} \equiv \omega_0 = -0.99\substack{+0.1 \\ -0.1}$ is consistent with the standard cosmological model i.e. with the cosmological constant. Note that around $z \approx 1$ the EoS changes concavity (inflexion point), producing a maximum in $d\omega/dz$.  Furthermore, the first derivative of $\omega$ with respect $z$ gives a value, as shown in figure~\ref{fig:param}, of $d\omega/dz\vert_{z=0}=-0.97\substack{+0.37 \\ -0.37}$, consistent with \cite{Riess:2004nr}.

%VM%
\subsection{Discriminating dark energy models}

The nature of DE is connected to the characteristics of the EoS. The reconstruction of the EoS by equation~\eqref{eq:wdef} may have singular points on its domain, i.e.  it might diverge, which occurs when the denominator is equal to zero. To find the singular points we consider the next equation: 
\begin{align}
\label{eq:sp1}
{1-\Omega_{m,0}(1+z)^3 \left({ H_0 \over {H(z)}} \right) ^2}=0\text{,}
\end{align}
\noindent which can be written as (see \ref{sec:A1}):
\begin{align}
\label{eq:sp2}
{1-\Omega_{m}(z)}=0\text{.}
\end{align} \\

\noindent We expect $\Omega_{m}(z)$ to be a monotonically increasing function from the present  (at $z=0$), to a matter dominated epoch when $q(z) \rightarrow q_1 = 1/2$ (see appendix~\ref{sec:A1}). As equation~\eqref{eq:q} asymptotically tends to  $q(z) \sim q_1$ as $z \rightarrow \infty$~\cite{de1970asymptotic},  our EoS reconstruction is valid only from today to an epoch of the Universe when matter dominates. In future works, we expect to study a more general parameterization of the deceleration by using $q_1$ as free parameter. %With the aim of finding the maximum value

%%%%%%%%%%%%%%%%%%%%%%%%%%%%%%%%%%% Image %%%%%%%%%%%%%%%%%%%%%%%%%%%%%%
%%%%%%%%%%%%%%%%%%%%%%%%%%%%%%
\begin{figure}\vspace{ 0.5cm}
\captionsetup{justification=centering}
\resizebox{0.49\textwidth}{!}{\includegraphics{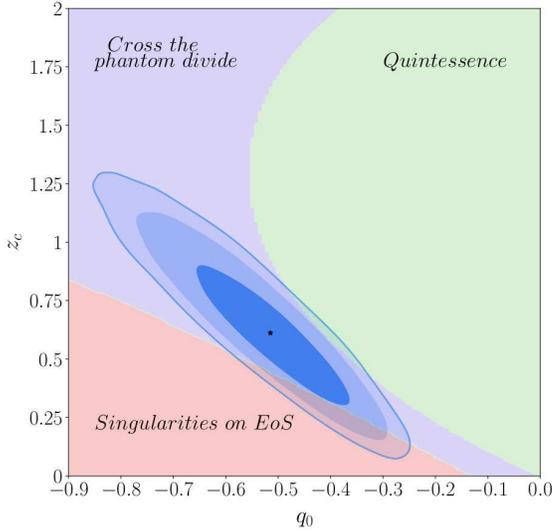}}
\captionof{figure}{
Decision regions in the parameter space and the $68\%$, $95\%$, $99.7\%$ confidence levels for the $q_{0}$ and $z_{c}$ parameters for the joint analysis constraints. The classification of the EoS (depending on the given   $\omega$ value) is represented in different regions: green for quintesence models; purple for quintom models; red  for EoS with singular points. The black star represents the joint analysis (CC + Pantheon) mean values for $z_c$ and $q_0$. \label{fig:dec_reg}}
\end{figure}\vspace{1cm}
%%%%%%%%%%%%%%%%%%%%%%%%%%%%%%%

The condition given by equation~\eqref{eq:sp2} is satisfied for $z > 0$. As comment before, the EoS is valid too in a matter dominated epoch, i.e., $z >> 1$,  let us assume for simplicity that $z \rightarrow \infty$. Thus, by substituting the equation~\eqref{eq:H} into equation~\eqref{eq:newomat}, the limit for $\Omega_{m}(z)$ at such epoch is:

\begin{align}
\label{eq:limo}
\lim_{z \to \infty} {\Omega_{m}(z)}=\Omega_{m,0}\ \text{exp}\left[2\xi(\erf(z_c)-1)\right] \text{.}
\end{align}

Considering that the reconstruction of $\Omega_m(z)$ for our model is a monotonic increasing function for $z\geq 0$ (see appendix~\ref{sec:A1}), for given a pair of fixed $q_0$ and $z_c$  there exist a real positive value of the redshift $z$ for which equation~\eqref{eq:wdef} will contain a singular point if 
\begin{align}
\label{eq:condeos}
\Omega_{m,0}\ \text{exp}\left[2\xi(\erf(z_c)-1)\right]\ >\ 1 \text{.}
\end{align}

Figure~\ref{fig:dec_reg} illustrates the $q_{0}-z_{c}$ region bounded for this inequality, showing two  regions of interest: the quintessence region and, where the EoS crosses the phantom divide. In the case that $\omega(z) \in [-1,1]$, the bartropic fluid can be represented with a minimally coupled scalar field, {{known as quintessence DE model and which is consistent with $\Lambda$CDM \cite{CruzDombriz2016}}}, but if $\omega(z) < -1$ the behavior of the fluid is represented as a phantom DE~\cite{Copeland:2006}. Since in our proposed EoS (see equation \eqref{eq:wdef}) does not exist an evident restriction for its codomain, it is important to know  whether the reconstruction go through the phantom divide, defined as $\omega = -1$. If the EoS cross the phantom divide, the DE behavior can be represented by {{the dynamics of more than a single scalar field \cite{vikman2005can}}, e.g.
a combination of a negative-kinetic and a normal scalar field, as quintom DE~\cite{Feng:2005}. Notice that our joint analysis mean values for $q_{0}$ and $z_{c}$  rely on both regions, panthom and quintessence.}

Quintessence models can be classified by the behavior of the potential associated with the scalar field. The two categories are thawing models and freezing (tracking) models (see ~\cite{Pantazis:2016,Sangwan:2018zpz} and references therein). In the thawing models, the scalar field is frozen at early times due to the Hubble parameter damping\footnote{ Indicates that the dynamics of the scalar field is governed by the Klein-Gordon equation},  while at late times the friction term becomes subdominant.  The $\omega(z)$ is a decreasing function that  asymptotically reaches the cosmological constant EoS (i.e. $\omega \approx -1$) at early times. In the freezing models, the scalar potential is steep enough at early times to develop the kinetic term, while at late times it becomes shallower allowing the slowing down of the scalar field. The $\omega(z)$ is an increasing function that 
tends to the Cosmological Constant EoS at late times. An effective tool to discriminate between these models is the $\omega'$-$\omega$ plane, where $\omega'\ =\ d\omega/dlna$~\cite{caldwell2005limits}; since different models are bounded by different regions~\cite{Chiba:2005tj,caldwell2005limits,Scherrer:2005je}.

\begin{figure}\vspace{ 0.0cm}
\captionsetup{justification=centering}
\resizebox{0.52\textwidth}{!}{\includegraphics{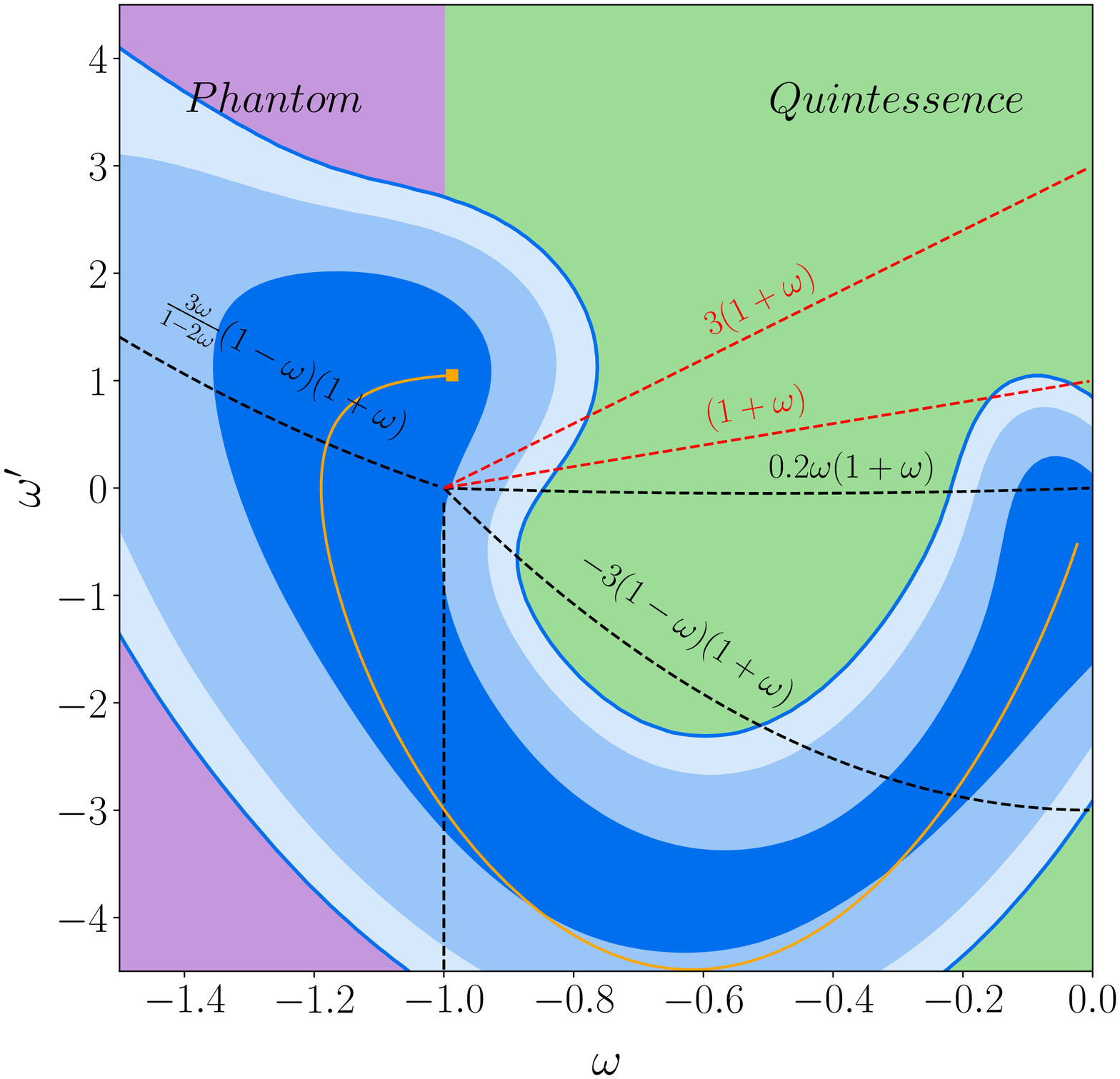}}
\captionof{figure}{Discrimination regions for quintessence (thawing and freezing behavior) and phantom models in the $\omega'$-$\omega$ plane. The red dashed lines represent the bounds for the thawing discrimination region, where $\omega'=3(1+\omega)$ is the upper bound  and $\omega'=(1+\omega)$ is the lower bound~\cite{caldwell2005limits}. The black dashed lines in the quintessence region ($\omega > -1$) are the bounds for freezing models, where $\omega'=0.2\omega(1+\omega)$ is the upper bound, and $\omega'=-3(1-\omega)(1+\omega)$ is the lower bound, see~\cite{Scherrer:2005je}. In the phantom region ($\omega < -1$), $\omega'=3\omega(1-\omega)(1+\omega)/(1-2\omega)$ is the upper bound, and $\omega=-1$ is the lower bound, see~\cite{Chiba:2005tj}. In shades of blue are the $68\%$, $95\%$, $99.7\%$ confidence levels for the reconstruction of $\omega$ and $\omega'$ using the joint constraints (CC + +Pantheon), and the orange line is the mean value of these reconstructions. The orange square is the value at redshift $z=0$.
}
 \label{fig:ToF}
\end{figure}

A phantom DE can be represented by a scalar field minimally coupled to gravity with a non-canonical negative-kinetic energy term, and whose energy density grows with time. Thus, the tracking behavior of a phantom model can be depicted in the $\omega'$-$\omega$ plane~\cite{Chiba:2005tj}. Because in the quintom models the evolution equations of the negative-kinetic and the normal scalar fields are independent~\cite{Guo:2004fq}, the potential behavior can be classified by the quintessence and phantom discrimination regions obtained separately. Figure \ref{fig:ToF} shows the discrimination regions for {quinte\-ssence} (thawing and freezing behavior) and phantom models in the $\omega'$-$\omega$ plane. The thawing discrimination region is delimited between $\omega'=1+\omega$ (lower bound) and $\omega'=3(1+\omega)$ (upper bound)~\cite{caldwell2005limits}. The freezing quintessence limits are provided by $\omega'=0.2\omega(1+\omega)$ (upper bound) and $\omega'=-3(1-\omega)(1+\omega)$ (lower bound)~\cite{Scherrer:2005je,Chiba:2005tj}. The upper bound for phantom region is $\omega'=3\omega(1-\omega)(1+\omega)/(1-2\omega)$~\cite{Chiba:2005tj}. {As shown in figure \ref{fig:ToF}, our analysis exclude thawing behaviour of the scalar field, being consistent with \cite{Dhawan:2017leu}.} Notice that our joint constraints on the $q(z)$ parameters crosses both the quintessence and phantom regions, hence, confirming that our results are consistent with {DE models that crosses the phantom divide, e.g. quintom DE}.

%%%%%%%%%%%%%%%%%%%%%%%%%%%%%%% Image %%%%%%%%%%%%%%%%%%%%%%%%%%%
%%%%%%%%%%%%%%%%%%%%%%%%%%%%%%%%%%%%%%%%%%%%%%
\vspace{0cm}
%%%%%%%%%%%%%%%%%%%%%%%%%%%%%%%%%%%%%%%%%%%%%%%%%

%%%%%%%%%%%%%%%%%%%%%%%%%%%%%%%%%%%%%%%%%%%%

\begin{table*}
\caption{ Mean values for the model parameters ($h$, $q_{0}$, $z_{c}$) derived from OHD and SN Ia measurements.}
\centering
\resizebox{1.0\textwidth}{!}{
\begin{tabular}{|lcccccccc|}
%\hline
\hline
Data set & $\chi^{2}_{min}$ &$h$& $q_{0}$& $z_{c}$ &$M_b^1$& $\delta_M$& $\alpha$& $\beta$ \\
\hline
\multicolumn{9}{|c|}{}\\
OHD (CC)  & $15$& $0.726^{+0.015}_{-0.015}$ & $-0.79^{+0.20}_{-0.14}$ & $0.81^{+0.22}_{-0.26}$ &---&---&---&---\\
& &  &  &  & & & & \\
SNIa (JLA) & $683$ & $0.722^{+0.19}_{-0.18}$ & $-0.44^{+0.12}_{-0.15}$ & $0.51^{+0.40}_{-0.30}$& $-18.98^{+0.50}_{-0.64}$ & $-0.07^{+0.02}_{-0.02}$& $0.14^{+0.006}_{-0.006}$&$3.10^{+0.08}_{-0.08}$ \\
& &  &  &  & & & & \\
SNIa (Pantheon) & $1035$ & $0.50^{+0.34}_{-0.34}$ & $-0.54^{+0.12}_{-0.14}$ & $0.75^{+0.32}_{-0.28}$& --- & ---& --- & --- \\
& &  &  &  & & & & \\
Joint (CC+JLA) & $700$ & $0.713^{+0.01}_{-0.01}$ & $-0.49^{+0.11}_{-0.12}$ & $0.51^{+0.22}_{-0.20}$& $-19.01^{+0.04}_{-0.04}$ & $-0.07^{+0.02}_{-0.02}$& $0.14^{+0.006}_{-0.006}$&$3.11^{+0.08}_{-0.08}$ \\
& &  &  &  & & & & \\
Joint (CC+Pantheon) & $1054$ & $0.710^{+0.01}_{-0.01}$ & $-0.51^{+0.09}_{-0.10}$ & $0.61^{+0.21}_{-0.18}$& --- & --- & ---  & --- \\
\hline
\end{tabular}}
\label{tab:par}
\end{table*}

%%%%%%%%%%%%%%%%%%%%%%%%%%%%%%%%%%%%%%%%%%%%

\section{Summary}
\label{sec:S6}

There are several ways to approach the dynamical evolution of the Universe with the aim of describing the late and early epoch expansion. Many models of DE, such as canonical and negative-kinetic scalar field models, are represented by a barotropic fluid. Recent observations indicate a transition between a decelerated and an accelerated phase of the cosmic expansion, from a matter dominated epoch to recent times, respectively. In this work we proposed a new phenomenological parameterization of the deceleration parameter, equation~\eqref{eq:q}, to approach the accelerated evolution of the cosmic expansion. The proposed form of $q(z)$ is a well behaved equation that can represent a step-like transition for this parameter, as well as being suitable for an analytical reconstruction of the Hubble parameter and the DE EoS. The behaviour of this new $q(z)$ parameterization allows to constrain minimally coupled scalar field DE models, as well as models which the DE EoS crosses the phantom divide. For minimally coupled scalar field models, as quintessence, the changes in the concavity of the proposed $q(z)$ points to a more general fitting of the {dynamics} of the scalar field, as thawing and freezing behaviours.

We performed an MCMC Bayesian analysis to constrain the $q(z)$ parameters using the OHD, and two SNIa samples: the JLA, and Pantheon. For the join analysis (CC + Pantheon) we obtain $q_0=-0.51\substack{+0.09 \\ -0.10}$, $h=0.710\substack{+0.01 \\ -0.01}$, and $z_t=0.65\substack{+0.19 \\ -0.17}$ , which are consistent with values reported by other authors. The reconstruction of the EoS (see figure~\ref{fig:dec_reg})  using these values crosses the phantom divide, rejecting the quintessence DE models. Our result points to a quintom DE, and it is consistent with a non parametric reconstruction of the EoS using the latest cosmological observations (see ref.~\cite{zhao2017dynamical}) within the range of validity of the equation~\eqref{eq:wdef}.

The  behavior of the two free-parameter reconstruction of the EoS (equation~\eqref{eq:A_w}) is a more general expression, including both the thawing or freezing scalar field models. Indeed, the functional form of $\omega$  does not impose an $\it{a priori}$ category of scalar field model for its entire domain. Furthermore, the discrimination analysis we presented in figure~\ref{fig:ToF} is also consistent with a quintom DE model. {{Quintom DE is only an example of a model that need the dynamics of more than a single scalar field to cross the phantom divide. Considering another set of models would imply that the energy-momentum tensor may deviate from the perfect-fluid form as those studied by \cite{deffayet2010imperfect}, which are related to Hordenski  gravity \cite{horndeski1974second}, and consistent with the recent GW observations \cite{PhysRevLett.119.251302}. Then  we  may  assume  that  these  models should  be  non  significant  deviation  from  the perfect-fluid form in order to remain the validity  of  eq.~\eqref{eq:wdef}. Another possibility is to invoke non-linear physics to explain the transition of the phantom divide with a single scalar field, as mentioned in \cite{vikman2005can}.}} The confidence contours for  $\omega'$ vs. $\omega$, are not subsets of a single model region within the regions delimited by thawing and freezing models. This is a complex behavior of our two free parameter reconstruction of the EoS, in contrast to the parameterizations analyzed in ref.~\cite{Pantazis:2016}  

In a future work, we plan to extend the study presented here, and analyze the  consequences of the cosmic expansion in a early epoch by setting $q_1$ as a free parameter, and its repercussions on the behavior of the effective EoS. {{Also to consider a more general set of imperfect DE models.}}\\

{The authors thank the anonymous referee for invaluable remarks and suggestions, that helped to improve the paper. The authors thank Luis Ure\~na for his thoughtful comments. This research has been carried out thanks to PROGRAMA UNAM-DGAPA-PAPIIT IA102517. J.M. acknowledges support from grant 3160674 (CONICYT /FONDECYT),}  and thanks the hospitality of the staff of IA-Ensenada
where part of this work was done.

\appendix

\section{The behavior of $\Omega_m(z)$  and the singularities of $\omega$}\label{sec:A1}

By considering the definition of the matter density in terms of $z$:
\begin{align}
\label{eq:omatdef}
\Omega_m(z)=\frac{\rho_m(z)}{3H^2(s)} \text{,}
\end{align}
where $\rho_m(z)=3H_0^2\Omega_{m,0}(1+z)^3$, equation~\eqref{eq:omatdef} can be rewritten as
\begin{align}
\label{eq:newomat}
\Omega_m(z)=\Omega_{m,0}\ (1+z)^3\ \left(\frac{H_0}{H(z)}\right)^2 \text{.}
\end{align}\\

Let us calculate the first derivative of $\Omega_m(z)$ with respect of $z$
\begin{dmath}
\label{eq:dom1}
\frac{d\Omega_m(z)}{dz}= 3\Omega_{m,0}\left(\frac{H_0}{H(z)}\right)^2(1+z)^2  - 2\Omega_{m,0}\frac{H_0^2}{H(z)^3} \frac{dH(z)}{dz}(1+z)^3  \text{,}
\end{dmath}
from equation~\eqref{eq:Hdef} $\frac{dH(z)}{dz}=H(z)\frac{1+q(z)}{1+z}$, simplifying equation~\eqref{eq:dom1}
\begin{align}
\label{eq:dom2}
\frac{d\Omega_m(z)}{dz}=\Omega_{m,0}(1+z)^2 \left(\frac{H_0}{H(z)}\right)^2(1-2q(z))  \text{,}
\end{align}
By the reconstruction of the Hubble parameter using the joint dataset, $H(z)>0$ and $q(z)<1/2$ for $z\geq0$, see figure~\ref{fig:hz_qz}. Introducing both considerations in equation~\eqref{eq:dom2}, we obtain
\begin{align}
\label{eq:dom3}
\frac{d\Omega_m(z)}{dz} > 0 \qquad  \forall z\geq 0 \text{,} 
\end{align}
for our model. Thus, in this case, $\Omega_m(z)$ is a monotonic increasing function for all $z\geq 0$. Given equation~\eqref{eq:limo} and $\Omega_m(0)=\Omega_{m,0}<1$ \cite{Riess:2004nr}, then the codomain of this function is delimited:
\begin{multline}
\label{eq:delomat}
{\Omega_{m}(z)} \in [\ \Omega_{m,0}\ ,\ \Omega_{m,0}\ \text{exp}(2\xi(\erf(z_c)-1))\ ) \cr \ \cr \forall z \geq 0 \text{.}
\end{multline}
Let us consider the next cases:\\
\begin{itemize}

\item If $\Omega_{m,0}\ \text{exp}(2\xi(\erf(z_c)-1))\ <\ 1$:
\begin{align}
\label{eq:C11}
\RA & \Omega_m(z)<1 & \forall z \geq 0 \\
\label{eq:C12}
\RA & 1-\Omega_m(z)>0 & \forall z \geq 0 
\end{align} 
then, there is not a value of $z \geq 0$ such that $\omega_\ef(z)$ diverges.\\

\item If $\Omega_{m,0}\ \text{exp}(2\xi(\erf(z_c)-1))\ =\ 1$:
\begin{align}
\label{eq:C21}
\RA & \Omega_m(z) \sim 1 & \text{as } z \rightarrow \inf\\
\label{eq:C22}
\RA & \omega_\ef(z) \rightarrow \inf & \text{as } z \rightarrow  \inf
\end{align}
then, $\omega_\ef(z)$ diverges as $z \rightarrow \inf$.

\item If  $\Omega_{m,0}\ \text{exp}(2\xi(\erf(z_c)-1))\ >\ 1$:\\

Because the codomain of $\Omega_m(z)$ is delimited as equation~\eqref{eq:delomat}, 
\begin{align}
\label{eq:1inomat}
1 \in [\ \Omega_{m,0}\ ,\ \Omega_{m,0}\ \text{exp}(2\xi(\erf(z_c)-1))\ ) \text{,}
\end{align}
then there is a value $z' > 0$ such that $\Omega_m(z')=1$
\begin{align}
\label{eq:C31}
\RA & 1-\Omega_m(z')=0  & \text{where } z' > 0.
\end{align}

\noindent Therefore, the last case gives the condition to have a  singular point of $\omega_\ef$.
\end{itemize}

%\begin{figure}
%\resizebox{0.75\textwidth}{!}{%\includegraphics{leer.eps}}
%\vspace{5cm}       % Give the correct figure height in cm
%\end{figure}
%

%

%
% BibTeX users please use
\bibliographystyle{unsrt}
\bibliography{article}
%\begin{thebibliography}{}

%Author, \textit{Book title} (Publisher, place year) page numbers
% etc
%\end{thebibliography}

\end{document}